\documentstyle[aps,preprint,tighten,epsfig,amssymb]{revtex}
\begin{document}
\title{
Probing meson spectral functions with double differential dilepton
spectra in heavy-ion collisions
}
\author{
B. K\"ampfer$^a$, O.P. Pavlenko$^{a,b}$
}
\address{
$^a$Forschungszentrum Rossendorf, PF 510119, 01314 Dresden, Germany \\
$^b$Institute for Theoretical Physics, 252143 Kiev - 143, Ukraine}

\maketitle

\begin{abstract}
The double differential dilepton spectrum
$dN/ (dM_\perp^2 \, dM^2)$ at fixed transverse mass $M_\perp$
allows a direct
access to the vector meson spectral functions. Within a fireball model
the sensitivity of $dN/ (dM_\perp^2 \, dM^2)$ against variations
of both the in-medium properties of mesons and the dynamics of the
fireball is investigated.
In contrast to the integrated invariant-mass spectrum
$dN/dM^2$, in the spectrum $dN/dM^2 \, dM_\perp^2$ with fixed $M_\perp$ 
the  $\omega$ signal is clearly seen as bump riding on the
$\rho$ background even in case of strong in-medium modifications.
\end{abstract}
\pacs{PACS number(s):25.75.-q, 25.75.Dw, 24.10.Cn}

\section{Introduction} 

A strong motivation for the starting experiments with the HADES detector
\cite{J.Friese}
at the heavy-ion synchrotron SIS in GSI Darmstadt is to search for
direct evidences of in-medium modifications of hadrons via the
dilepton ($e^+ e^-$) channel.
Such modifications, like the $\rho$ meson ''melting'' and possible
in-medium mass shifts, being related to chiral symmetry restoration,
are considered now as the main cause of the low-mass dilepton
excess observed in heavy-ion experiments at CERN-SPS energies
\cite{Rapp_Wambach}. According to the present understanding
of the chiral phase transition \cite{Phys.Rep.,Pisarski},
the chiral condensate $\langle q \bar q \rangle$ as order parameter
decreases with increasing temperature $T$ and baryon density $n$.
This leads to modifications of the meson spectral functions
since they are coupled to $\langle q \bar q \rangle$.
While the dependence of the chiral condensate as a function of the
temperature at small baryon density is quite smooth up to a critical
temperature of $T_c \sim 170$ MeV \cite{Karsch}, the dropping of
$\langle q \bar q \rangle$ with increasing density 
at small temperature is more pronounced
already at moderate densities. For instance, in the QCD sum rule
analysis \cite{Hatsuda}, which is in accordance with the
anticipated Brown-Rho scaling \cite{BR},
a density dependence of the hadron mass shift like
$m^* = m (1 - 0.18 n/n_0)$ has been predicted, so that
even at the nuclear saturation density 
$n = n_0 = 0.15$ fm${}^{-3}$ a noticeable change
of the in-medium mass $m^*$ emerges.
Unfortunately, in the case of heavy-ion collisions, where densities
up to $3 n_0$ can be expected at SIS energies, the remarkable advantage
of the strong density dependence meets a problem when trying to identify
the predicted mass shift by the direct dilepton decay of vector
mesons. 
Due to the
collective expansion of the baryonic matter the density $n$ depends on
time so that obviously a time average of the mass shift is to be
expected, i.e. the mass shift effect is smeared out over an interval
related to the change of $n$.

To extract the wanted information on the in-medium meson spectral
function via dilepton measurements in heavy-ion collisions
one has also to take into account the general properties of the
dilepton spectra in a thermalized system which undergoes a
collective expansion. For instance, in a dense system not too far from
local thermal equilibrium, the dilepton rate is known to be governed
mainly by the Boltzmann factor $\exp \{-E / T \}$, where
$E = \sqrt{M^2 + \vec Q^2}$ is the lepton pair energy, and the pair
is characterized by the invariant mass $M$ and three-momentum $\vec Q$.
That means the dilepton spectrum $dN/dM^2$ as a function of the
invariant mass is a convolution of spectral functions and the
Boltzmann factor. For temperatures $T < 100$ MeV, as estimated for
SIS energies, and dilepton masses in the region of the light vector
mesons, $M \sim m_V \sim 770$ MeV, the ratio $M/T$ can be in the order
of 10 and, as a consequence, the resulting spectra are very steep.
This can make an identification of possible mass shifts, superimposed
to (at least collision) broadening, rather difficult. Moreover,
dynamical effects, caused by flow and cooling, can even stronger mask 
the wanted information.

In the present note we propose to extend the usual analysis of the
integrated spectrum $dN/dM^2$ to the more informative double differential
spectrum $dN/(dM_\perp^2 \, dM^2$), where $M_\perp = \sqrt{M^2 + Q_\perp^2}$
is the transverse mass of the pair with transverse two-momentum
$\vec Q_\perp$. This allows to avoid the majority of the above mentioned
difficulties in extracting the in-medium vector meson spectral
function on the basis of dilepton measurements in heavy-ion collisions.
In particular, by taking the transverse mass $M_\perp$ at a fixed value
(or in a sufficiently narrow bin) one gets rid of the steep slope of the
invariant mass spectrum as we show below. As a result the spectral function
is seen more clearly, and possible in-medium effects become more
pronounced.

To demonstrate the very idea of our approach we employ here a fireball
model for describing the space-time evolution of the dense matter
formed in central collisions of heavy nuclei at SIS energies. In spite of the
schematic character of the model, it is useful to estimate various
characteristics, such as the relative contributions of vector mesons
decaying inside and outside of the fireball 
and Dalitz decays of mesons and the
$\Delta$ as well, which are related to observables accessible in 
HADES experiments.
Along with more complicated transport model calculations 
\cite{transport,Frankfurt},
we consider our approach as a first step towards elucidating the most
appropriate observables for verifying in-medium changes of meson
properties in heavy-ion collisions.

Our paper is organized as follows. In Sect.\ 2, we present the 
parameterizations of the dilepton rates which we use in our
exploratory studies. The fireball dynamics is given in  Sect.\ 3.
The resulting dilepton spectra are discussed in Sect.\ 4,
where we contrast the dileptons emitted during the fireball's
lifetime and such dileptons which stem from hadron decays
after freeze-out.
The discussion can be found in sect.\ 5 together with remarks
on the support of the presented results by calculations based
on a BUU code.  

\section{Dilepton rates} 

The dilepton rate from vector meson decays,
$V \to e^+ e^-$ with $V = \rho, \omega, \phi$,
in an ideal resonance gas approximation is given by
\begin{equation}
\frac{dN}{d^4x \, d^4Q} = \sum_V
\frac{2 d_V}{(2 \pi)^3} \exp \left\{- \frac{u \cdot Q}{T} \right\}
\frac 1 \pi
\frac{(m_V \Gamma_V^{\rm tot})(M \Gamma_{V \to e^+ e^-})}
{(M^2 -m_V^2)^2 +(m_V \Gamma_V^{\rm tot})^2},
\label{eq.1}
\end{equation}
where $\Gamma_V^{\rm tot}$ and $\Gamma_{V \to e^+ e^-}$ denote the
total and dilepton decay widths, $m_V$ is the vector meson mass,
and $d_V$ stands for the degeneracy factor.
The lepton pair has the four-momentum
$Q^\mu = (M_\perp \cosh Y, M_\perp \sinh Y, \vec Q_\perp)$,
where $Y$ denotes the pair rapidity.
In (\ref{eq.1}) we use the Breit-Wigner parameterization 
of the meson spectral function,
\begin{equation}
A_V(m) =
\frac 1 \pi
\frac{m_V \Gamma_V^{\rm tot}}
{(m^2 -m_V^2)^2 +(m_V \Gamma_V^{\rm tot})^2},
\label{eq.2}
\end{equation}
which can be thought to result from the generalization of an 
on-mass shell delta function to the case of an unstable particle
\cite{P.Koch}. Such an ansatz is also used for QCD sum rule
analyzes \cite{Leupold} of vector meson properties in a nuclear
medium. Due to four-momentum conservation one gets immediately
$m^2 = Q^2 \equiv M^2$.

It should be emphasized that, within the vector dominance model,
the dependences
$\Gamma_{\rho \to e^+ e^-} \propto M^{-2}$
and
$\Gamma_\rho^{\rm tot} \propto M^2$ just cancel, therefore,
in (\ref{eq.1}) the widths are understood to be constant.
A remaining $M$ dependence stemming from   
$\Gamma_{\rho \to \pi^+ \pi^-}^{\rm tot} \propto (1- \frac{4
m_\pi^2}{M^2})^{3/2}$
is partially mimicked by the factor $M$ in the numerator in (\ref{eq.1}).
We have checked that the rate described by (\ref{eq.1}) agrees
with the standard parameterization \cite{P.Koch}
of the process $\pi^+ \pi^- \to e^+ e^-$
in the intervals 400 MeV\footnote{At this and smaller values of $M$
the rate is anyhow dominated by Dalitz decays.
Therefore, we do not attempt a modification of the low-$M$ tail of
(\protect\ref{eq.1})
which arises from a particular model. For more involved
expressions of the rate we refer the interested reader, e.g., to
\cite{C.Gale}, where also the coupling to the baryon charge is 
included.} $< M <$ 1000 MeV and
50 MeV $< T <$ 100 MeV. For the narrow $\omega$ and $\phi$ mesons these 
subtleties are not important.

The collective expansion of the matter is characterized by the
four-velocity $u^\mu(x)$ depending generally on space and time, as
also does the temperature $T(x)$. In spite of the approximative character
of (\ref{eq.1}) it is useful for demonstrating how the
in-medium modifications in the spectral function (\ref{eq.2})
show up in the dilepton spectra if flow effects of the matter are
taken into account.

Integrating the rate (\ref{eq.1}) over the three-momentum of the lepton
pair one gets the rate as a function of the invariant mass,
\begin{equation}
\frac{dN}{d^4x \, dM^2} = \sum_V
\frac{d_V}{2 \pi^3} 
M T K_1 \left( \frac{M}{T} \right)
\frac{(m_V \Gamma_V^{\rm tot})(M \Gamma_{V \to e^+ e^-})}
{(M^2 -m_V^2)^2 +(m_V \Gamma_V^{\rm tot})^2},
\label{eq.3}
\end{equation}
where $K_1$ is  a Bessel functions.

In Fig.~1a we plot the rate (\ref{eq.3}) at a temperature 
of $T = 70$ MeV
to demonstrate the general tendency of the very steeply dropping
spectrum with increasing $M$.
Since we do not advocate in the present note any particular functional
form
of the meson spectral function (2), which is under intense debate
so far, we parameterize schematically here the in-medium width
broadening by the change of the total vacuum width
according to
$\Gamma_V \to \Gamma_V^* = b \Gamma_V$ with a constant parameter
b = 3 for all vector mesons.
(The position of the in-medium vector meson peak is not important 
for this example.)
As seen in Fig.~1a the important effect of the vector meson broadening
is smeared out by the rapidly dropping thermal weight factor
$K_1(M/T)$ in (\ref{eq.3}). No peak related to the $\rho$ meson
remains. The $\omega$ peak is still visible, but not very pronounced.
The strongest peak structure is related to the $\phi$ meson.

The above example can be contrasted to the double differential rate
$dN /(d^4 x dM_\perp^2 dM^2)$ obtained from (\ref{eq.1}) by
integrating over the dilepton rapidity and the polar angle. For a
spherically symmetric source we find
\begin{equation}
\frac{dN}{d^4x \, dM_\perp^2 \, dM^2} = \sum_V
\frac{2 d_V}{(2 \pi)^3} 
K_0 \left( \frac{M_\perp \gamma}{T} \right)
I_0 \left( \frac{Q_\perp v_r \gamma}{T} \right)
\frac{(m_V \Gamma_V^{\rm tot})(M \Gamma_{V \to e^+ e^-})}
{(M^2 -m_V^2)^2 +(m_V \Gamma_V^{\rm tot})^2},
\label{eq.4}
\end{equation}
where $\gamma = 1/\sqrt{1 - v_r^2}$ is the Lorentz factor
corresponding to the radial flow three-velocity $v_r$, and
$Q_\perp = \sqrt{M_\perp^2 - M^2}$.
One can infer from (\ref{eq.4}) that, by fixing the transverse mass,
the thermal weight is frozen in. For sufficiently small values of the flow
velocity $v_r$ the rate $dN/(d^4 x \, dM_\perp^2 \, dM^2)$ as function of
the invariant mass at fixed $M_\perp$ reflects directly the shape of the
meson spectral function. In Fig.~1b
we show this remarkable property
for $M_\perp = 1.3$ GeV, $v_r = 0.3$ and $T = 70$ MeV.
The $\rho$ bump is here clearly visible below the $\omega$ peak;
the $\phi$ needle is more pronounced. 
The price is, of course, a diminished count rate.
However, for $M_\perp = 0.9$ GeV the rate is still not too much smaller
in comparison with the rate displayed in
in Fig.~1a at $M = 0.9$ GeV.

In Fig.~1b we show that the rate increases in the $\rho$ region
with increasing values of the radial velocity $v_r$. This fact is known
since some time \cite{Kajantie} as enhancement of the $\rho$ peak due to
transverse flow. Note that the rate $dN/(d^4x dM^2)$ is completely
unaffected by flow effects.

It should be stressed that other representations of the double
differential rate, such as $dN/d^4 x \, dQ_\perp^2 \, dM^2$,
are less suitable for an access to the spectral function.

\section{Dynamics} 

To obtain the dilepton spectrum from the meson decay rate (\ref{eq.1})
one needs to specify the space-time evolution of the matter.
According to our experience \cite{Phys.Lett.} the detailed knowledge
of the space-time evolution is not necessary to describe the experimental
dilepton spectra in relativistic heavy-ion collisions
in the low-mass region, where the in-medium
modifications of the meson spectral function play a crucial role.
In this line, and also to avoid too many parameters, we employ here 
a variant of the
blast wave model \cite{blast_wave} with constant radial expansion
velocity $v_r$. In such a model the radius of the fireball increases
like $R(t) = R_i + v_r t$. From the analysis
of hadron ratios in heavy-ion reactions at SIS energies a freeze-out
temperature of $T_f = 50$ MeV at freeze-out baryon density
$n_f = 0.3 n_0$ is obtained \cite{Oeschler}.
Extrapolating back along an isentropic line by using a resonance gas model
as in \cite{Oeschler} up to $n_i = 3 n_0$ (as suggested by transport code
simulations for the maximum density at SIS energies) 
one finds $T_i = 90$ MeV. The freeze-out time $t_f$
is determined by baryon conservation within the fireball
volume; 
we use as baryon participant number $N = 330$. The time evolution
starts at initial time $t_i =0$.
We take $v_r = 0.3$ as a reasonable average according to
\cite{Oeschler}.
By these parameters the dynamics of the fireball is completely
fixed.

\section{Dilepton spectra} 

\subsection{The fireball radiation} 

Since the aim of our work is a demonstration of the advantage of the
double differential dilepton spectrum $dN/(dM_\perp^2 \, dM^2)$
independently of a given form of the vector meson's spectral function
$A_V$, we do not use specific functional changes of $A_V$ 
but consider a variety of
possible parameterizations of the in-medium effects via mass shifts
and width broadenings, i.e.,
changes of the parameters $\Gamma_V^{\rm tot}$ and $m_V$. 
As long as the quantitative description of
$A_V(m)$ in a medium is matter of debate \cite{Rapp_Wambach} such a
procedure seems appropriate.

We begin with an extreme case where the $\rho$ meson mass shift
is $\Delta m_\rho = 300$ MeV, while the $\omega$ meson is unaffected,
i.e. $\Delta m_\omega = 0$, and $m_V^* = m_V - \Delta m_V$.
The in-medium
widths of both the low-lying vector mesons is increased by $b = 3$.
The resulting spectra are displayed in Fig.~2, and we have
chosen $M_\perp = 0.9$ GeV. In contrast to the integrated rate
$dN/dM^2$, where the $\rho$ peak disappears (cf.\ Fig.~2a), 
one can clearly
observe the splitting of both vector meson peaks
in the $dN/(dM_\perp^2 \, dM^2)$ spectrum (cf.\ Fig.~2b), 
thus reflecting the original
modifications of the $\rho, \omega$ spectral functions.

Another similarly extreme case is to keep the $\rho$ mass fixed and to
shift the $\omega$ mass by
$\Delta m_\omega = 300$ MeV down; again with widths broadening
factor $b = 3$. As seen in Fig.~3 the information on the
change of the meson properties is best visible in the
$dN/(dM_\perp^2 \, dM^2)$ spectrum, where a strong $\omega$ peak and
some part of the $\rho$ bump are distinguishable (cf.\ Fig~3b). 
Otherwise, the shifted
$\omega$ peak is also nicely visible in the $dN/dM^2$ spectrum 
(cf.\ Fig.~3a) because
it ''rides'' on the $\rho$ background and becomes more pronounced;
a $\rho$ bump is here again hardly visible.

In more realistic considerations the values of the meson mass shifts
and width broadenings should depend on the baryon density and temperature
which are strongly time dependent. This in turn causes an additional
smearing of the peaks of the spectral functions. To demonstrate that
$dN/(dM_\perp^2 \, dM^2)$ is still useful to elucidate occurrence
of in-medium modifications 
we parameterize now the in-medium $\omega$ mass by
$m_\omega^* = m_\omega (1 - \delta_\omega \cdot n/n_0)$,
$\Gamma_\omega^{\rm tot * }= \Gamma_\omega^{\rm tot}
+ \delta \Gamma_\omega n/n_0$ with $\delta_\omega = 0.08$
and $\delta \Gamma_\omega = 20$ MeV,
and for the $\rho$ meson we assume only a broadening according to
$\Gamma_\rho^{\rm tot * }= \Gamma_\rho^{\rm tot} 
+ \delta \Gamma_\rho \cdot n/n_0$
with $\delta \Gamma_\rho = 400$ MeV,
where the superscript * denotes again the in-medium quantities.
These parameterizations are in line with current expectations
\cite{Klingl_Weise} that the $\rho$ meson suffers a very strong in-medium
broadening, while for the $\omega$ meson one can expect moderate
modifications. The resulting spectra are displayed in Fig.~4.
In contrast to the integrated invariant mass spectrum
$dN/dM^2$ (cf.\ Fig.~4a), in the spectrum
$dN/(dM_\perp^2 \, dM^2)$ with fixed $M_\perp$ (cf.\ Fig.~4b)
the $\omega$ meson contribution is clearly seen as shifted bump
riding on the $\rho$ background,
in spite of the smearing caused by the space time evolution.
(Notice that in Fig.~4a the smeared $\omega$ contribution
could easily be misinterpreted as $\rho$ bump.)
Therefore, the measurement of the double differential dilepton
spectrum at fixed $M_\perp$ offers much better chances to identify
a possible $\omega$ mass shift.

It should be stressed that changes of $T_i$ cause a noticeable
up/down shift of the spectra but let the shapes nearly the same.

\subsection{Background contributions}

Within the given model one can also estimate various background contributions.
For instance, the contributions of vector meson decays after freeze-out
can be consistently evaluated. The invariant mass spectrum of such decays
can be calculated from
\begin{equation}
\frac{dN}{dM^2} =
\sum_V
\frac{1}{\Gamma_V^{\rm tot}}
\frac{d \Gamma_{V \to e^+ e^-}}{dM^2}
N_V(T_f,m_V = M),
\label{eq.5}
\end{equation}
where $d \Gamma_{V \to e^+ e^-}/dM^2 = \Gamma_{V \to e^+ e^-} A_V(M^2)$
and $N_V(T_f,m_V)$ is the number of vector mesons at freeze-out.
This number is calculable within the Cooper-Frye formalism
\cite{Cooper_Frye}, which includes relativistic effects
related to the matter's expansion. In our spherical fireball model
with constant radial flow velocity one gets
\begin{eqnarray}
N_V (T_f, m_V) & = &
\frac{d_V}{(2 \pi)^3}
\frac{4 \pi R_f^3}{3 \gamma}
\int d m^2
\frac{m_V \Gamma_V^{\rm tot}}{(m^2 -m_V^2)^2 +(m_V \Gamma_V^{\rm
tot})^2}
\label{eq.6} \\
& \times &
\int d p_\perp^2
\sqrt{ \frac{2 \pi T_f}{\gamma m_\perp} }
\exp \left\{ - \frac{m_\perp \gamma}{T_f} \right\}
\left[ \frac{\sinh a_f}{a_f} (\gamma p_\perp + T_f)
- T_f \cosh a_f \right], \nonumber
\end{eqnarray}
where $m_\perp = \sqrt{m^2 + p_\perp^2}$ is the transverse mass
of the meson $V$ with transverse momentum $p_\perp$, and
$ a_f = \gamma v_r p_\perp /T_f$. In deriving (\ref{eq.6}) we employ
the meson spectral function $A_V(m)$ from (\ref{eq.2}).

The calculations show that
even for the ''long-living'' $\omega, \phi$ mesons the decay
contribution after freeze-out
is considerably smaller than the decay contribution during the 
fireball's life time. This is in
contrast to naive expectations based on simple life time arguments,
according to which the life time of the $\omega$ meson is 23 fm/c
compared to the life time of the thermalized system
${\cal O}(10)$ fm/c. To arrive at a consistent estimate one has to
take into account the evolution and the hypothesis of thermalization.
The latter assumption is the key to understand the small contribution
of the decays after freeze-out: 
In the early stage the temperature is much higher
and, according to the assumed chemical equilibrium, the appearance
of vector mesons is much more frequent than at freeze-out.
At the pole positions the ratio of the fireball to the freeze-out
decay contributions can be roughly estimated by
$\exp \{ - m_V/T_i \} / \exp \{ - m_V/T_f \}$, because of the 
dominating exponential factors. 

It should be emphasized
that the analysis of the hadron multiplicities at freeze-out
in \cite{Oeschler}
supports the chemical equilibrium picture for some hadrons.
In the present work we assume chemical equilibrium for the full
time evolution.
This equilibrium is certainly not maintained during 
the time evolution for all
hadron species. For instance, the $\omega$ production via the
inverse decay reaction $3 \pi \to \omega$ is rather unlikely.
Such effects can be simulated by additional chemical off-equilibrium
potentials, which keep certain hadron species on a higher number
than we have assumed here. In this respect our estimates,
e.g., of the $\omega$ numbers, serve as unfavorable lower limits
when assuming particle production in the stages of highest
temperature and density.

While we focus on the direct decay channels $V \to e^+ e^-$,
the Dalitz decay channels, like 
$\Delta \to N e^+ e^-$,
$\phi \to \eta e^+ e^-$, 
$\omega \to \pi^0 e^+ e^-$,
$\eta'\to \pi^0 (\eta) e^+ e^-$, 
$\eta \to \gamma e^+ e^-$,
$\pi_0 \to \gamma e^+ e^-$ etc.,
populate the invariant mass spectrum in the low-mass region
(for a recent comprehensive analysis of the various other channels cf.\
\cite{Tuebingen}).
These Dalitz decays are 
calculated by (5, 6)
with $d \Gamma_{V \to e^+ e^-}/dM^2$ to be replaced by
$d \Gamma_{A \to B e^+ e^-}/dM^2$ (for hadron decays after freeze-out)
and  
\begin{eqnarray}
\frac{dN_{A \to B e^+ e^-}}{d^4x dM^2} & = &
\frac{d_A}{2 \pi^3} \int dm m^2
\frac{m_A \Gamma_A^{\rm tot}}
{(m^2 - m_A^2)^2 + (m_A \Gamma_A^{\rm tot})^2}
\frac{d \Gamma_{A \to B e^+ e^-}}{dM^2}
T K_1 \left( \frac{m}{T} \right), \\
\frac{d \Gamma_{A \to B e^+ e^-}}{dM^2} & = &
\frac{\alpha}{3 \pi M^2} \Gamma_{A \to B \gamma^*},
\end{eqnarray}
(for hadron decays during the fireball's life time ), 
where $\Gamma_{A \to B \gamma^*}$
is taken from \cite{P.Koch} for meson decays and from
\cite{Frankfurt} for $\Delta$ decays.

With respect to the $\Delta$ Dalitz contribution we mention that
the number of $\Delta$ resonances depends on the baryo-chemical
potential. The latter one is in our fireball model approximately
850 MeV and fairly well time independent. The quoted value results
from the chemical freeze-out analysis in \cite{Oeschler}, and
the approximate time independence is a special feature of the
isentropic line with specific entropy of 5 
in the interval $n = 0.3 \cdots 3 n_0$.

We have checked that the sum of these Dalitz decay channels
at $M > 0.5$ GeV is below the direct vector meson decay
contributions (see also remarks in the next section). 

\section{Discussion and summary}

In summary we have demonstrated that, within a fireball model
and a simple parameterization of the rate,
the double differential dilepton spectra
$dN/(dM_\perp^2 \, dM^2)$ at fixed value of $M_\perp$
can deliver valuable information on a possible change of the
in-medium properties of vector mesons. The usually considered
invariant mass spectrum $dN/dM^2$ contains a convolution
of the spectral function with a thermal weighting function which
is too strong such that
modifications of the spectral functions are difficult to disentangle,
in particular if the dynamics of the system causes an additional
smearing.

We consider our fireball model as first step towards finding the most
appropriate characteristics of dilepton spectra to probe in-medium
vector meson spectral functions. The results obtained, in particular
the $dN/dM^2 \, dM_\perp^2$ spectra, should be confirmed by more
realistic transport calculations. Indeed, using the BUU code developed
from \cite{Gyuri_old} we performed calculations in line with the 
present work. These calculations show very clearly that, when
selecting the interval 850 MeV $< M_\perp <$ 950 MeV,
(i) the resulting spectrum  $dN/dM^2 \, dM_\perp^2$ is dominated
by the channel $\pi^+ \pi^- \to \rho \to e^+ e^-$ in the interval
300 MeV $< M <$ 850 MeV thus directly reflecting the shape of the 
$\rho$ formfactor,
(ii) in the high-$M$ tail also direct $\rho$ decays, with $\rho$'s
stemming from other channels than $\pi^+ \pi^-$ annihilations, can
become as large as the $\pi^+ \pi^-$ annihilation contribution, 
(iii) the $\omega$ rides on the $\rho$ bump, and
(iv) the $pn$ bremsstrahlung, $\Delta$ Dalitz and $\eta$ Dalitz
decays are negligible in the mentioned $M$ interval.
The details of these BUU calculations will be presented elsewhere
\cite{Wolf}.

To make reliable predictions for the future experiments at
HADES one needs also additional analyses of the mass shifts and widths
with respect to both the baryon density and temperature dependences
in the regions expected to be achieved in heavy-ion collisions
at SIS.

{\bf Acknowledgments:} 
We thank M. Gorenstein, 
R. Rapp, 
A. Sibirtsev, 
Gy. Wolf and 
G. Zinovjev
for stimulating discussions.
O.P.P. thanks for the warm hospitality of the nuclear theory group
in the Research Center Rossendorf.
The work is supported by BMBF grant 06DR829/1, WTZ UKR-008/98, and
STCU-015.



\begin{figure}[h]
\centering
~\\[-.1cm]
\psfig{file=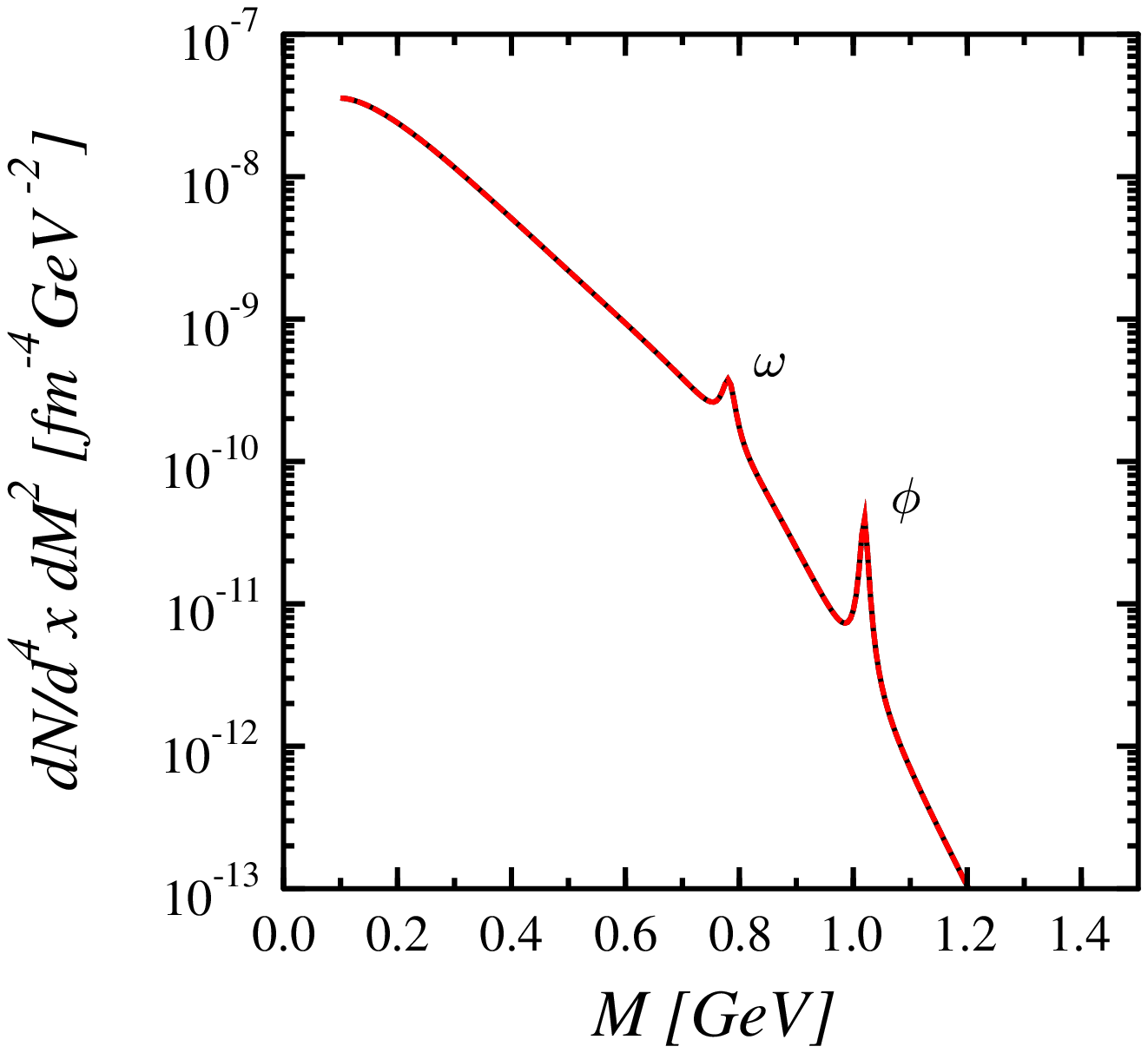,width=7.5cm,angle=-0}
\hfill
\psfig{file=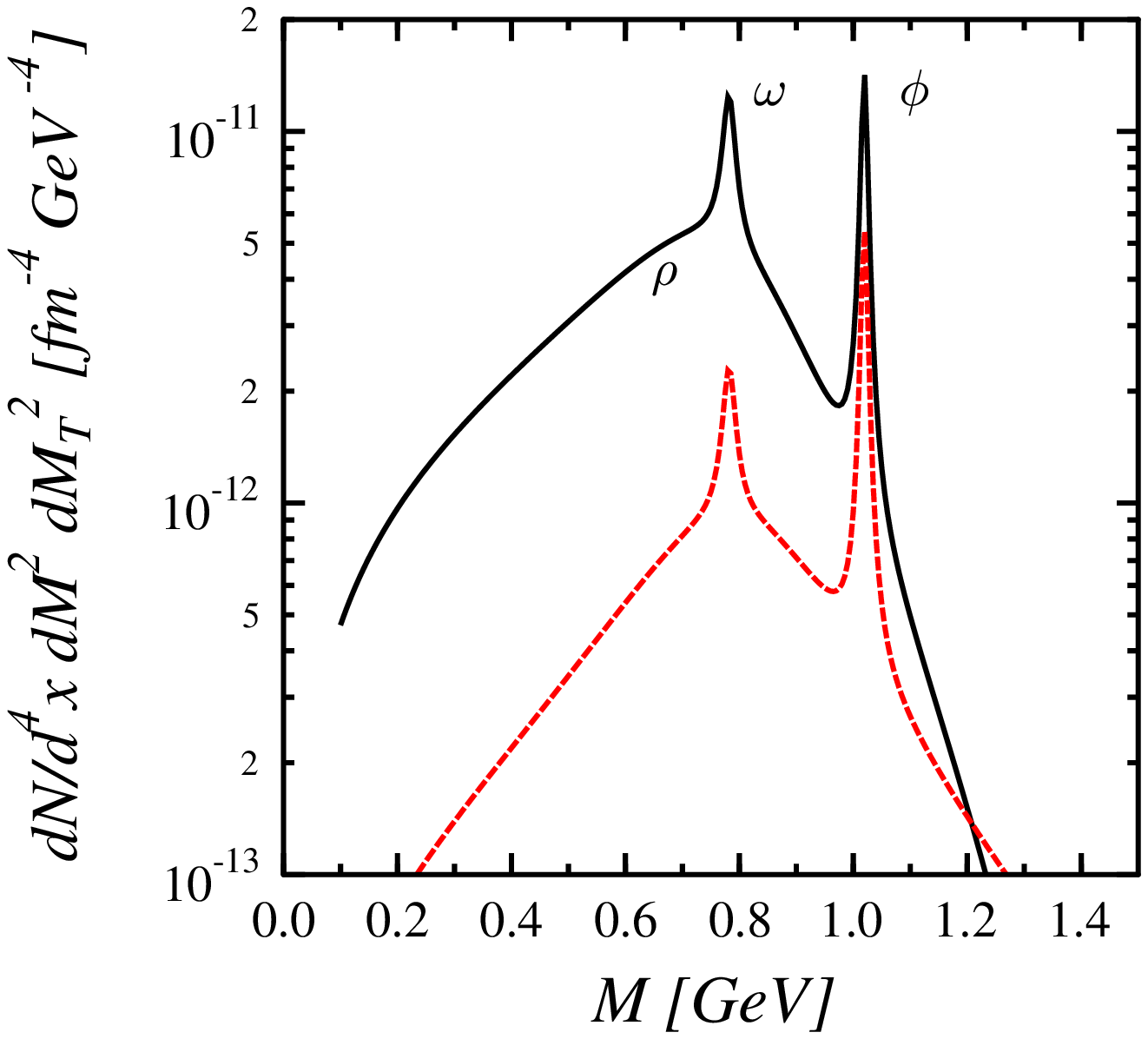,width=7.5cm,angle=-0}
~\\[.1cm]
\caption{
The rates $dN/(d^4 x \, dM^2)$ ((a) left panel) and
$dN/(d^4 x \, dM_\perp^2 \, dM^2)$ ((b) right panel, $M_\perp = 1.3$ GeV)
for $T = 70$ MeV, $v_r = 0.1$ and 0.3, and $b = 3$.
}
\label{fig.1}
\end{figure}

\begin{figure}[h]
\centering
~\\[-.1cm]
\psfig{file=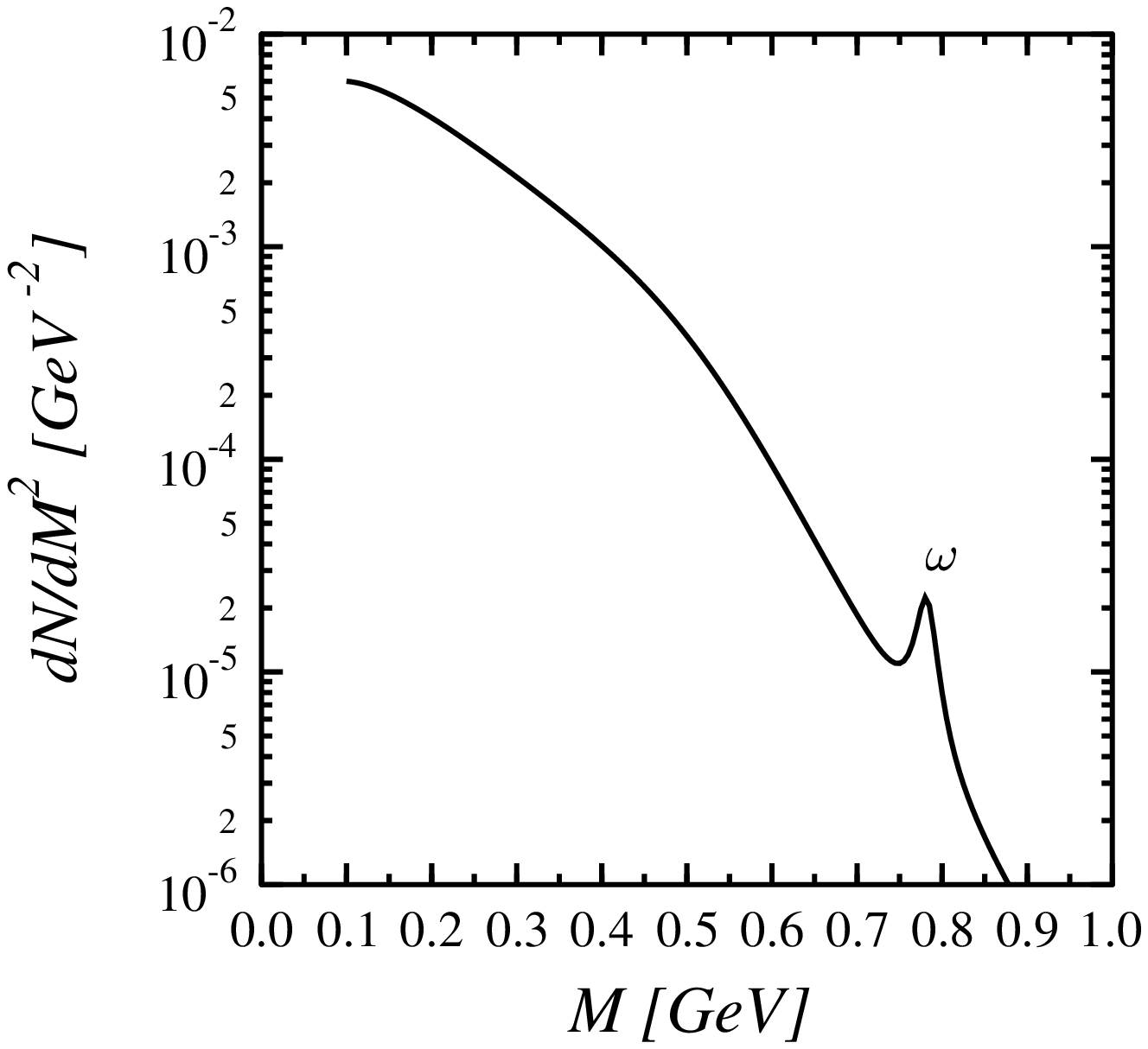,width=7.5cm,angle=-0}
\hfill
\psfig{file=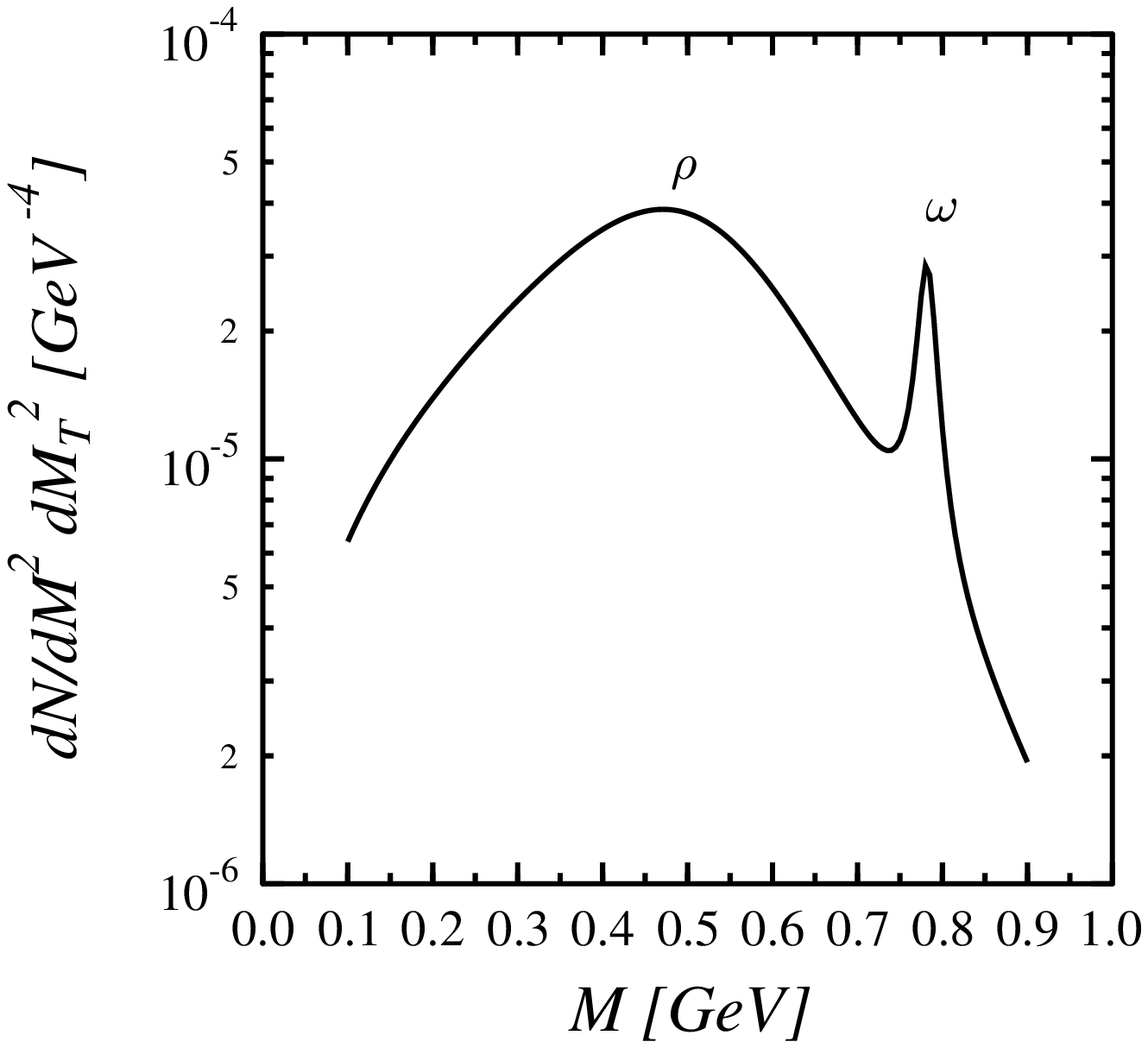,width=7.5cm,angle=-0}
~\\[.1cm]
\caption{
The rates $dN/(dM^2)$ ((a) left panel) and
$dN/(dM_\perp^2 \, dM^2)$ ((b) right panel, $M_\perp = 0.9$ GeV)
for the dynamical scenario described in the text and for
$b = 3$, $\Delta m_\rho = 300$ MeV,  $\Delta m_\omega = 0$.
}
\label{fig.2}
\end{figure}

\begin{figure}[h]
\centering
~\\[-.1cm]
\psfig{file=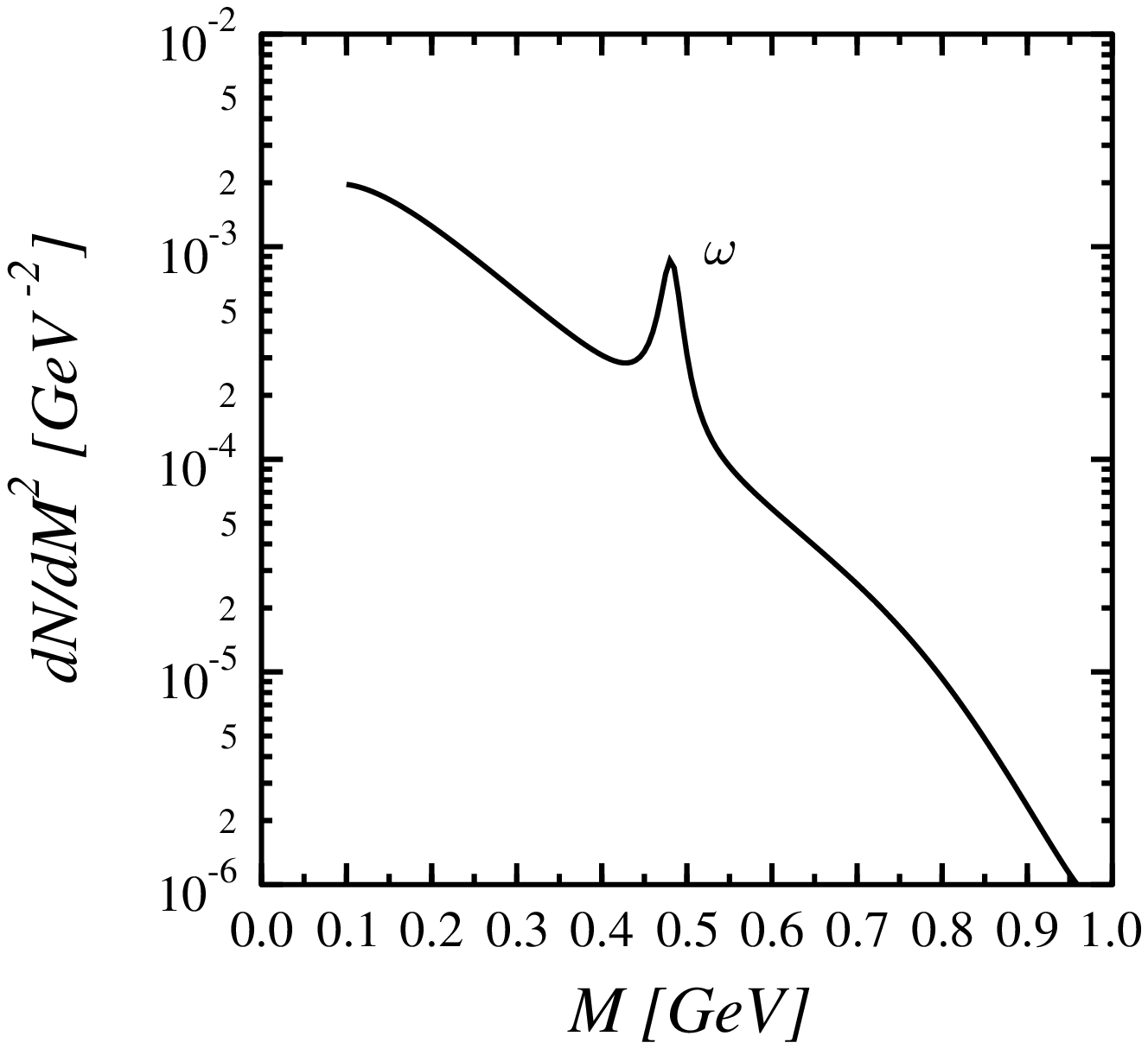,width=7.5cm,angle=-0}
\hfill
\psfig{file=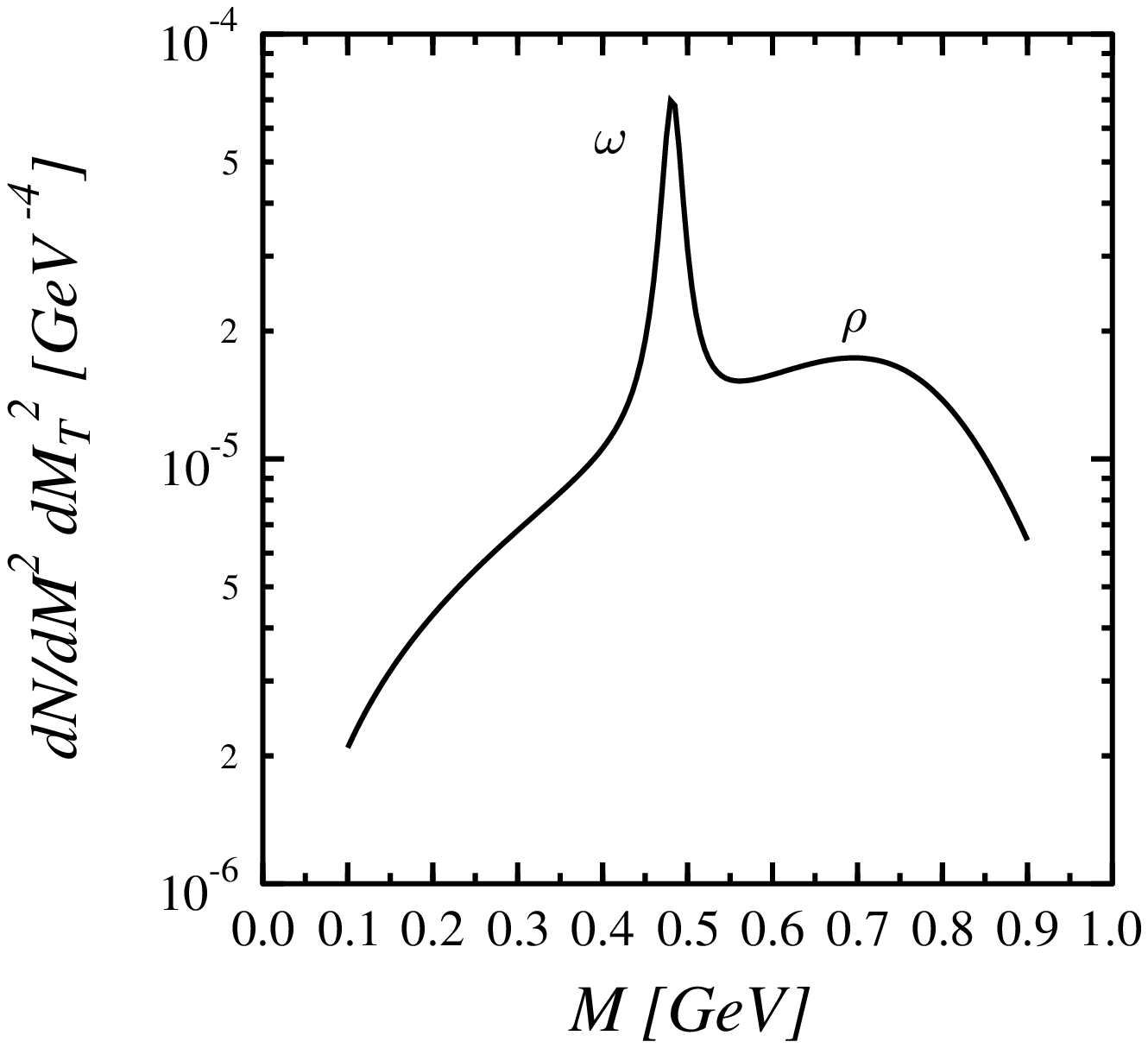,width=7.5cm,angle=-0}
~\\[.1cm]
\caption{
As in Fig.~\protect\ref{fig.2} but for
$\Delta m_\rho = 0$,  $\Delta m_\omega = 300$ MeV.
}
\label{fig.3}
\end{figure}

\begin{figure}[h]
\centering
~\\[-.1cm]
\psfig{file=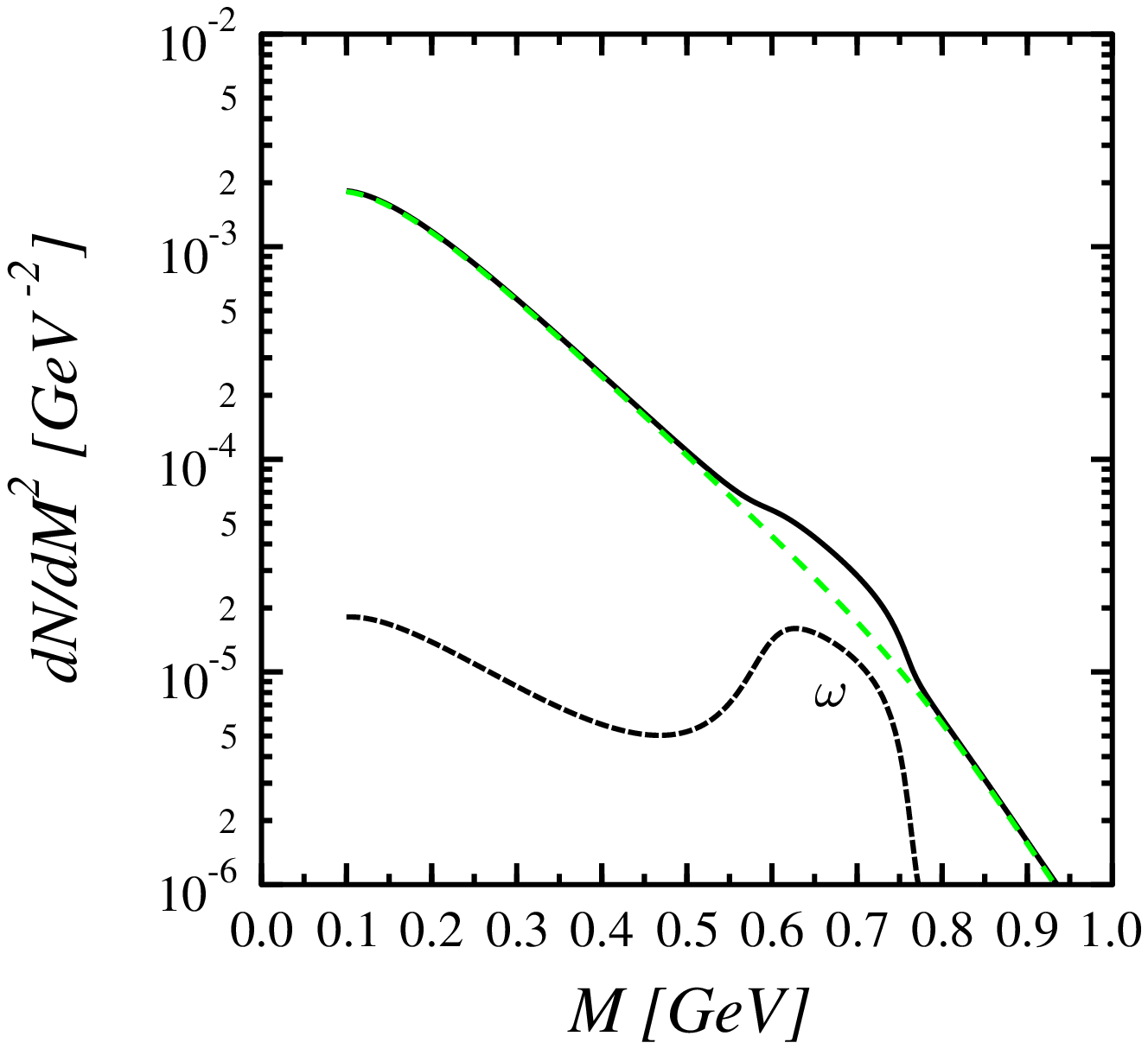,width=7.5cm,angle=-0}
\hfill
\psfig{file=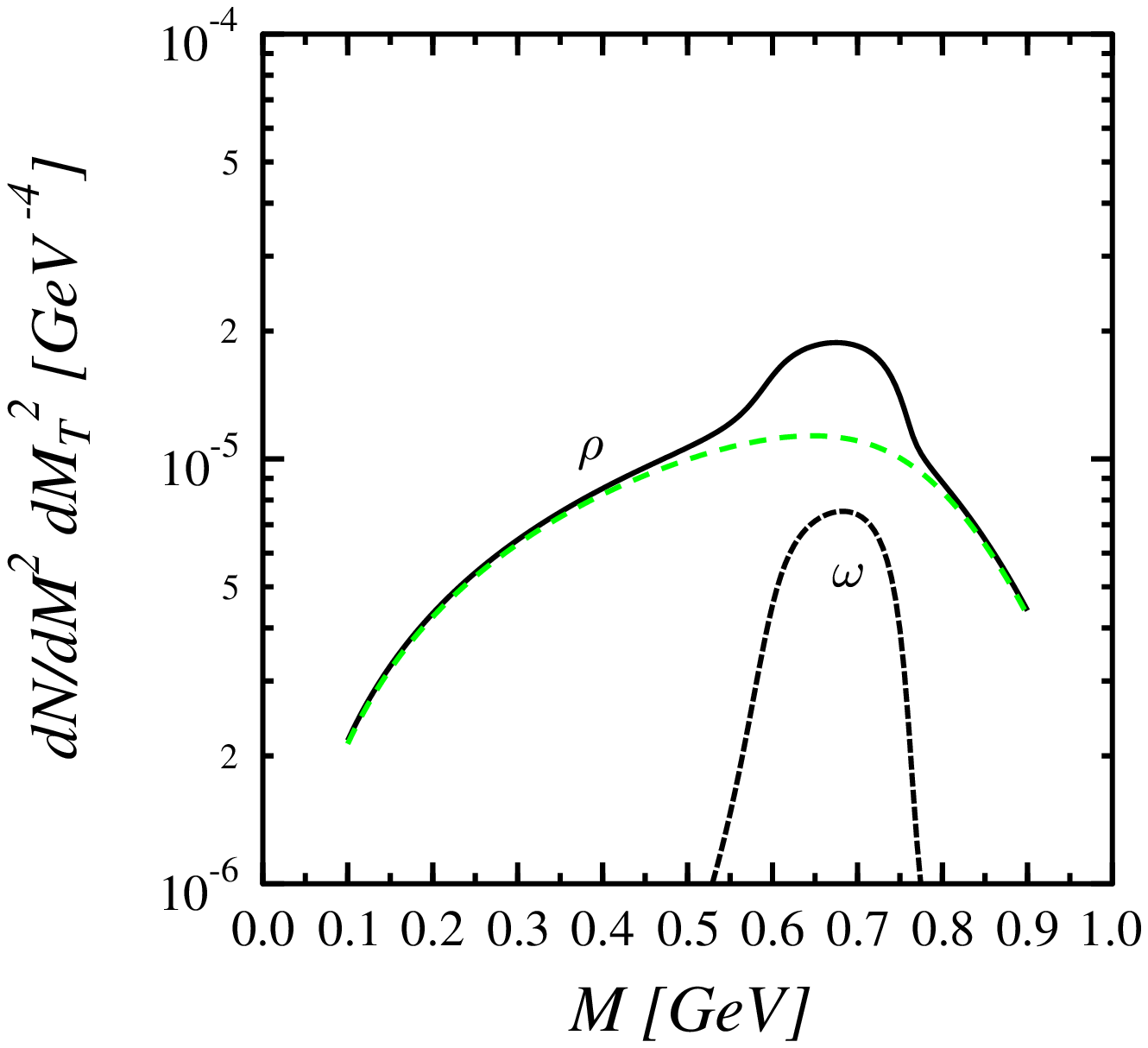,width=7.5cm,angle=-0}
~\\[.1cm]
\caption{
As in Fig.~\protect\ref{fig.2} but for
$\delta_\omega = 0.08$,
$\delta \Gamma_\omega = 20$ MeV,
$\delta_\rho = 0$,
$\delta \Gamma_\rho = 400$ MeV.
The $\omega$ ($\rho$) contribution is displayed separately as
short-dashed (long-dashed) line. 
}
\label{fig.4}
\end{figure}

\end{document}